# Nonlinear Terahertz Electroluminescence from Dirac–Landau Polaritons

B. Benhamou-Bui[1], C. Consejo[1], S.S. Krishtopenko[1], S. Ruffenach[1], C. Bray[1], J. Torres[1], J. Dzian[2], F. Le Mardelé[2], A. Pagot[3], X. Baudry[3], S.V. Morozov[4,5,6], N.N. Mikhailov[7,8], S.A. Dvoretskii[8,9], B. Jouault[1], P. Ballet[3], M. Orlita[2], C. Ciuti[10], F. Teppe[1]

[1] *Laboratoire Charles Coulomb (L2C), UMR 5221 CNRS – Université de Montpellier, F-34095 Montpellier, France*

[2] *Laboratoire National des Champs Magnétiques Intenses, CNRS – UGA – UPS – INSA – EMFL, Grenoble, France*

[3] *CEA, LETI, MINATEC Campus, DOPT, Grenoble, France*

[4] *Institute for Physics of Microstructures of the Russian Academy of Sciences, Nizhny Novgorod, Russia*

[5] *Lobachevsky State University of Nizhny Novgorod, Nizhny Novgorod, Russia*

[6] *Ioffe Institute, 194021 St. Petersburg, Russia*

[7] *A.V. Rzhanov Institute of Semiconductor Physics, Siberian Branch of the Russian Academy of Sciences, Novosibirsk, Russia*

[8] *Novosibirsk State University, Novosibirsk, Russia*

[9] *Tomsk State University, Tomsk, Russia and*

[10] *Université Paris Cité, CNRS, Matériaux et Phénomènes Quantiques, 75013 Paris, France*

We report Dirac–Landau polaritons observed by terahertz (THz) magnetoreflectivity spectroscopy, demonstrating strong coupling between cyclotron transitions of two-dimensional (2D) Dirac fermions in HgTe quantum wells and optical cavity modes. Under pulsed electrical injection we observe efficient nonlinear electroluminescence, with a strongly out-of-equilibrium polariton distribution dominated by emission from the upper polariton branches. Model analysis of the bias-dependent emission intensity and spectral narrowing indicates a polariton occupancy per mode approaching unity, with a possible contribution from stimulated polariton emission in the spectral region of the upper anticrossing. These results open prospects toward Dirac–Landau polariton condensates and low-threshold, tunable THz polariton lasers based on cyclotron emission.

*Introduction* — Dirac materials [1] are attractive for cyclotron-based light emission because their non-equidistant Landau level spectra suppress Auger recombination and can enable cyclotron emission [2–4]. Their small carrier mass also allows wide tunability at low magnetic fields [5, 6], unlike p-Ge Landau level lasers where the heavy hole mass required much larger fields [7–11]. In graphene, however, equidistant Landau-level subsets remain and sustain residual Auger channels [6]. HgTe quantum wells (QWs) near the topological phase transition [12–14] instead exhibit quadratic corrections to the Dirac spectrum that strongly reduce near-equidistant spacings, creating especially favorable conditions for intense Landau emission, as shown by recent spontaneous cyclotron emission measurements [15, 16]. These QWs thus provide a powerful platform to study cyclotron emission and non-equilibrium carrier dynamics. Yet stimulated emission is still difficult, since population inversion typically requires electric fields close to the material breakdown limit.

In parallel, strong light-matter coupling enables hybrid bosonic quasiparticles, polaritons, with quantum-fluid behavior [17] and condensation/lasing without electronic population inversion [18, 19]. Exciton-polariton lasers span many material platforms [20–26], but a polariton laser based on inter-Landau level transitions has not been achieved. Landau polaritons—formed by coupling Landau transitions to confined photonic modes—were predicted [27] and observed by reflectivity/transmission in conventional semiconductor 2DEGs [28–30]. More recently, new regimes and platforms have been explored [31–36]. Combining Dirac Landau levels with strong coupling thus offers a new route toward THz polariton condensates and lasers.

In this Letter, we observe Dirac–Landau polaritons in HgTe QWs arising from strong coupling between cyclotron transitions of two-dimensional Dirac fermions and THz cavity modes. Magneto-reflectivity reveals clear anticrossings between cavity modes and Landau transitions, confirming hybrid light–matter excitations. We then demonstrate polaritonic electroluminescence under pulsed electrical excitation in crossed electric and magnetic fields. Emission peaks follow the polariton branches extracted from reflectivity, and the nonlinear bias dependence together with spectral narrowing indicates a polariton occupancy per mode close to unity with a possible contribution from stimulated polariton emission.

*Dirac-Landau polaritons* — The polariton excitations of a cavity-embedded two-dimensional electron gas in a magnetic field can be described by a bosonic quantum Hamiltonian [27] of the form $H = H_{cavity} + H_{Landau} + H_{int} + H_{dia}$, where $H_{cavity}$ is the bare cavity Hamiltonian, $H_{Landau}$ describes the collective cyclotron excitation of Landau-quantized electrons, $H_{int}$ is the paramagnetic light-matter interaction, and $H_{dia}$ is the diamagnetic term [27]. This effective description involves only bosonic operators for the cavity photons and the cyclotron excitation mode. The matter component of a Landau polariton is a collective cyclotron excitation, first established in parabolic GaAs/AlGaAs quantum wells [28]. Collective cyclotron excitations also occur for non-parabolic dispersions, as observed in Ref. 32, and in Dirac systems with linear bands. Graphene cavity QED [37–39] has been explored theoretically in the ultrastrong-coupling regime, where possible ground-state instabilities depend on the role of the diamagnetic term in effective theories. In ultrastrong coupling [40], when the coupling is comparable to cavity and cyclotron frequencies, diamagnetic effects become central. Here, instead, we focus on excited states (polaritons) in the strong-coupling regime, where diamagnetic corrections are moderate, and we emphasize the nonlinear emission of electrically driven Dirac–Landau polaritons.

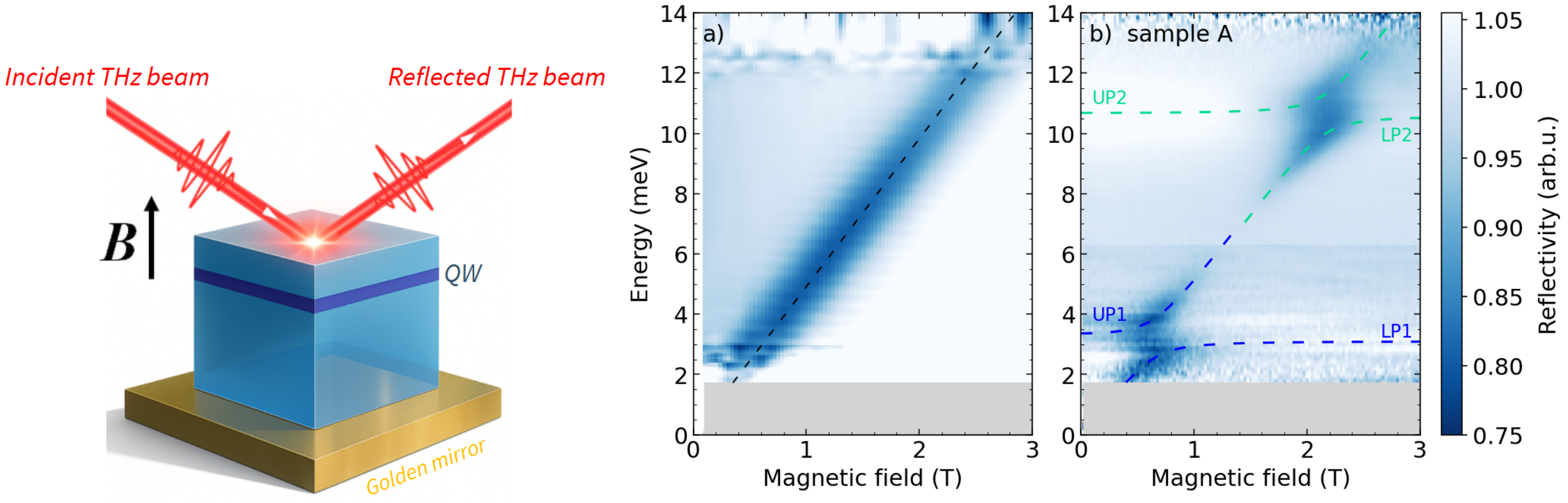

FIG. 1. Left panel: Sketch of the physical system, consisting of a 2D material hosting Dirac fermions embedded in a THz cavity resonator. A magnetic field $B$ is applied perpendicular to the plane of the 2D material, inducing Landau quantization of the electronic states. The optical response of the system is probed via reflection spectroscopy using an incident THz beam. Right panel: Magneto-reflectivity spectra (a) without a cavity and (b) with a cavity resonator. In the absence of a cavity, the color plot shows a linear evolution of the cyclotron energy with increasing magnetic field. The 2D material in Sample A is an 8 nm-thick HgTe quantum well measured at $T = 4.2$ K. In the presence of a 28 $\mu$m-thick cavity, the cyclotron dispersion displays clear spectral anti-crossings at magnetic fields of 0.7 T and 2 T, corresponding to energies of approximately 3 meV and 10 meV, respectively.

To study cyclotron emission from Dirac–Landau polaritons, we first realize strong coupling between a resonant THz cavity and the cyclotron resonance of Dirac fermions. We performed THz magneto-absorption spectroscopy (Fig. 1). The system is an 8-nm-thick HgTe/CdTe QW, close to the gapless state and grown by molecular beam epitaxy, integrated into a THz optical cavity (see Figs. 1 and 2). The THz Fabry–Perot cavity is formed by thinning the substrate (GaAs for samples A, B and D, CdTe for sample C) to enhance photon confinement and light–matter coupling. A gold coating on the substrate back side forms the mirror, while the semiconductor–helium interface provides the front partially reflective boundary. Further details are given in the Supplemental Material (SM). Before magneto-reflectivity, we used two characterization methods. First, we measured multiple substrates with thicknesses from 300 $\mu$m to 40 $\mu$m by THz time-domain spectroscopy (TDS). Second, we measured the same samples in emitter mode by detecting their thermal radiation at zero magnetic field under electrical heating [41–45]. Both methods confirm an optical cavity within the substrates (SM). We then analyzed the narrowest cavity by relative magneto-reflectivity to extract the coupling between cyclotron resonance and cavity modes, and compared to identical measurements without a THz cavity (Fig. 1a,b).

Without the cavity, magneto-reflectivity shows a cyclotron resonance whose energy evolves linearly with magnetic field. The system lies in the Shubnikov-de Haas (incipient Landau quantization) regime, where Landau levels exist but partially overlap (SM). While the cyclotron energy is linear in $B$, its slope depends on the electron density. A linear fit yields a cyclotron mass 0.024 times the bare electron mass, consistent with the measured density. With the cavity, the cyclotron dispersion is strongly reshaped by coupling to two cavity modes with photon energies $\hbar\omega_1 \simeq 3$ meV and $\hbar\omega_2 \simeq 10$ meV. This produces pronounced avoided crossings [46]), yielding two pairs of upper/lower Landau polariton branches: (UP1, LP1) near $B = 0.7$ T and (UP2, LP2) near 2 T, demonstrating strong light–matter coupling. Hopfield-model fits for mixed light–matter quasiparticles in the strong-coupling regime give excellent agreement, with coupling strengths $\hbar\Omega_1 = 0.65$ meV and $\hbar\Omega_2 = 0.62$ meV.

*Electroluminescence* – We investigated cyclotron emission from Dirac electrons in HgTe QWs under quantizing magnetic fields by applying short electric pulses that populate high Landau levels with non-equilibrium carriers. Without a cavity, electroluminescence shows a Gaussian peak shifting linearly with $B$, matching the cyclotron-resonance slope from magneto-reflectivity (Figs. 2b and 1a). A weak low-energy tail appears at large $B$; enhanced by normalization (each spectrum scaled to its maximum), it originates from a small thermal background present in all samples (SM). With the cavity, the spectra change qualitatively (Figs. 2c–e): emission is suppressed below the first cavity mode and no longer follows a linear $B$ dependence. At low fields, discrete peaks at fixed energies correspond to cavity modes, reproducibly observed across samples and cavity thicknesses (SM). The 50 $\mu$m cavity displays three modes near 4, 7, and 10 meV (Fig. 2c), whereas the 28 $\mu$m cavity shows only two modes at 4 and 10 meV (Fig. 2d), with stronger coupling. At higher fields the peaks shift and anticross, signaling polariton formation. In Fig. 2e, peak energies from the 28 $\mu$m cavity show two inflections at 4 and 10 meV, matching UP1 and UP2 from reflectivity. These features also appear in Fig. 2f, a false-color map of sample A versus energy and magnetic field: the upper polariton branch is well resolved for the first mode, and a corresponding feature is also visible for the second. The Hopfield polariton model reproduces these data using the same parameters as magneto-reflectivity. We can additionally tune the polaritonic anticrossings with an electrostatic back gate that adjusts the carrier density in sample C (SM).

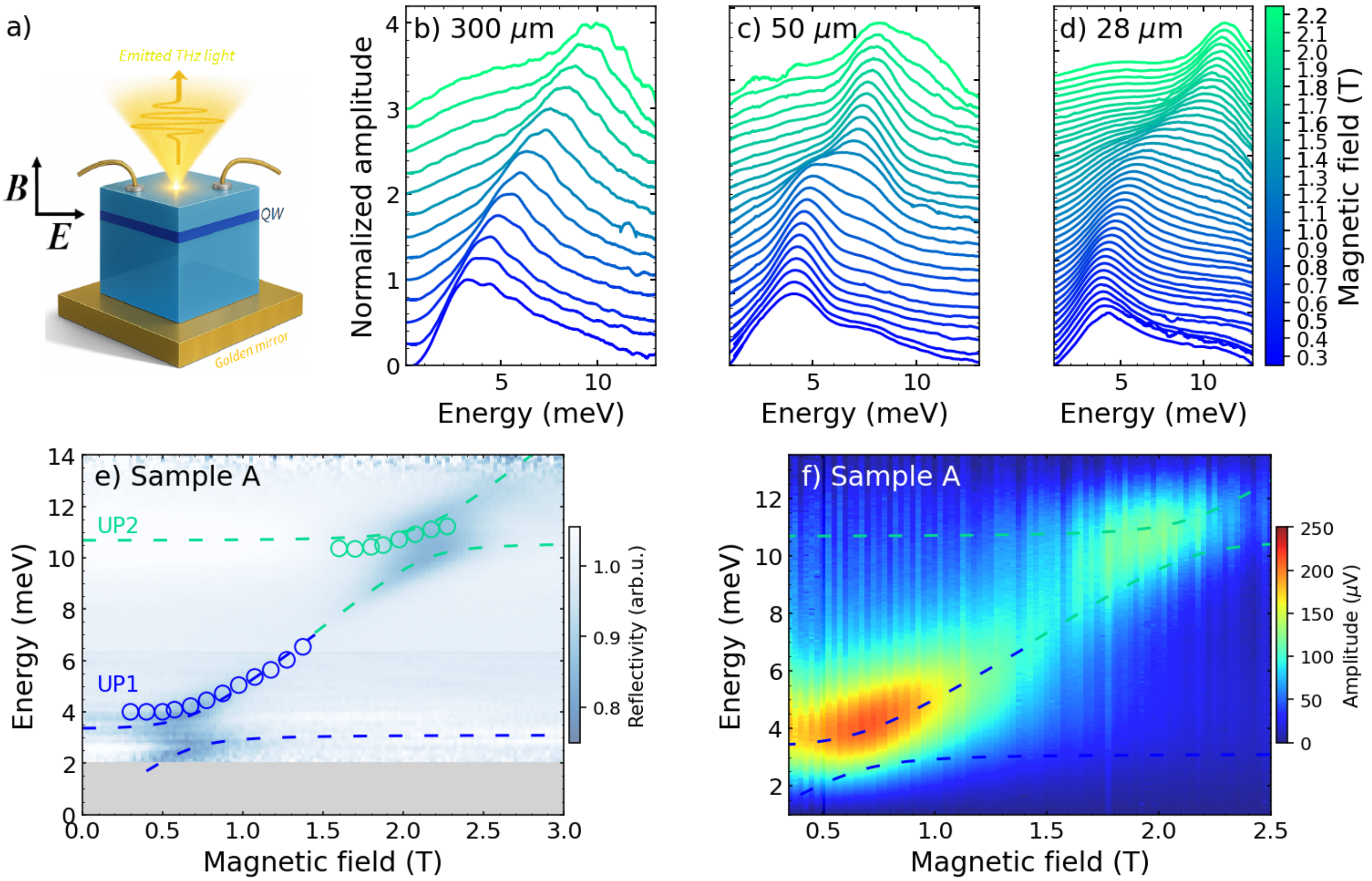

FIG. 2. (a) Sketch of the THz electroluminescent cavity device, where an in-plane electrical bias injects current between source and drain ohmic contacts. THz emission spectra at various magnetic fields, shown for different device configurations. (b) Without a cavity, on a 300 $\mu$m-thick substrate, the emission energy scales linearly with magnetic field. (c) With a 50 $\mu$m-thick cavity, three cavity modes appear around the energies 4, 7, and 10 meV, showing clear coupling to the cyclotron resonance. (d) With a 28 $\mu$m-thick cavity (sample A), only two modes at 4 and 10 meV are observed, exhibiting stronger line shifts. (e) The emission peak energies from panel (d), plotted versus magnetic field, align with the reflectivity features labeled UP1 and UP2 in Fig. 1b. (f) False-color map of the emission spectra from sample A as a function of energy and magnetic field. The dashed lines indicate the previously identified Hopfield-like fits.

The electroluminescence linewidth is broader than in absorption, as expected because we integrate over emission angles. Further broadening can arise from hot-carrier distributions [11], electron–phonon scattering [47], and Stark broadening by ionized impurities [48, 49]. Still, peak positions track the upper polariton dispersion (see SM), confirming emission from Dirac–Landau polaritons. Remarkably, only the upper branch is seen in emission. Thermal black-body radiation is excluded because it would favor lower energies. The strong non-equilibrium distribution may reflect a bottleneck [50–52] that suppresses relaxation toward lower-energy states. Unlike exciton-polaritons [21], where luminescence is dominated by the lower branch and bot-tleneck effects concern higher energies, here emission favors the upper branches. A possible origin is the reduced Landau-level spacing at high energies, which enhances losses for lower polariton states. Crossed electric and magnetic fields can further compress high-energy spacings [53], and Coulomb interactions may enable resonant conversion of polaritons into electron–hole excitations, as discussed in the "Emission / transmission comparison" section of the SM. Clarifying these relaxation channels will require further experiments and theory, in HgTe and other Dirac materials.

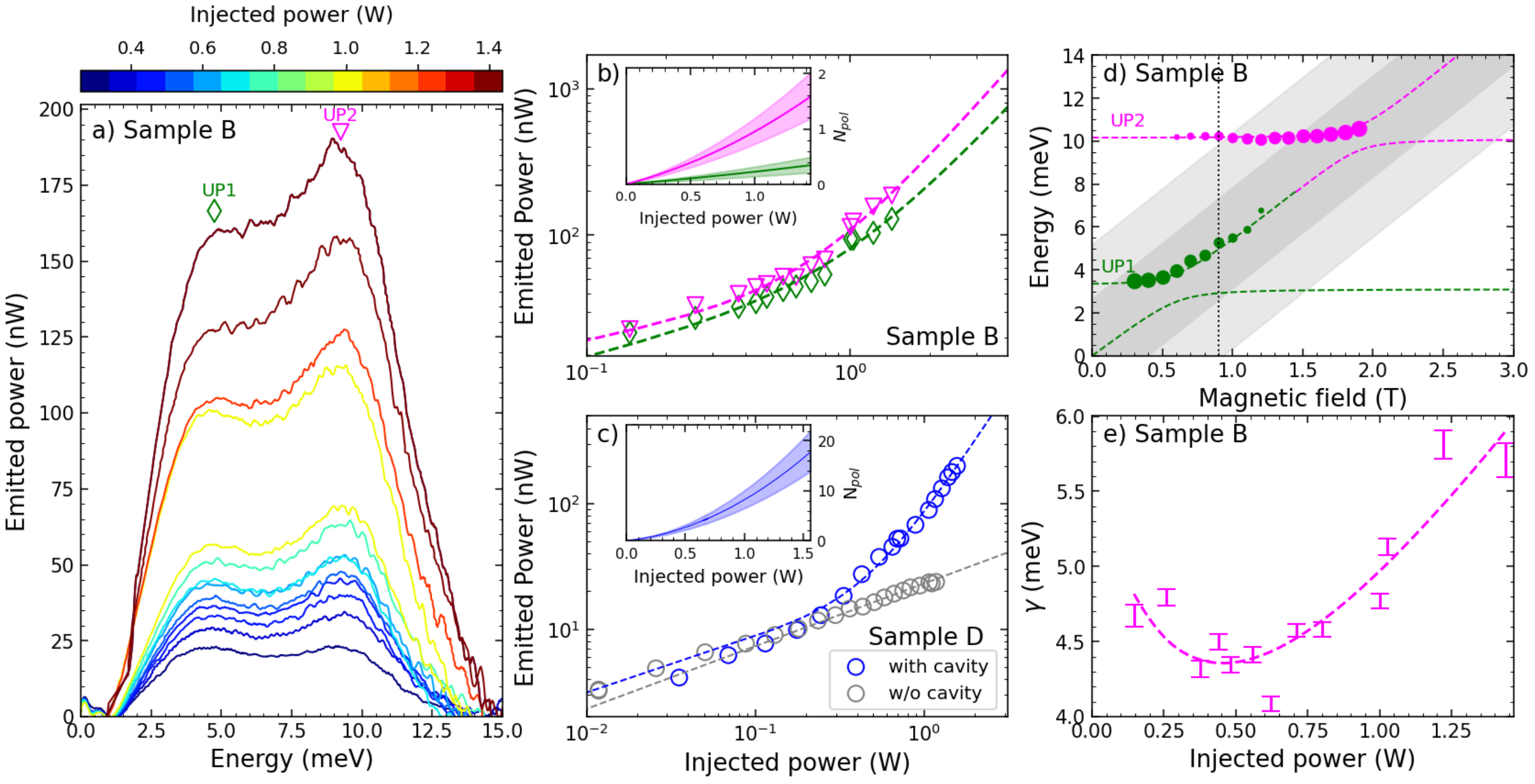

FIG. 3. (a) Electroluminescence spectra measured at $B = 0.9$ T on sample B for different values of injected electrical power from 0.15 to 1.45 W (corresponding to electric fields ranging from 300 to 1000 V/cm). (b) Evolution of the two emission peaks, corresponding to the UP1 and UP2 polariton modes, as a function of the injected electrical power. Marker styles and colors correspond to the peaks shown in (a). Dashed lines represent fits used to extract the polariton population evolution using Eq. S19 of the SM, shown in the inset. (c) Comparison of the emitted power versus the injected electrical power for the bare (gray points) and cavity (blue points) configurations of sample D at a magnetic field of 0.9 T. Dashed lines represent fits. The raw emission spectra from which these data points were extracted are shown in Supplementary Material (SM). The inset shows the dependence of the number of polaritons per mode on the electrical power. (d) Green and pink points marking the emission maxima, with their sizes reflecting the relative intensities of the two peaks. The dashed curves correspond to the Hopfield model. The shaded regions indicate the emission strength: strong (dark gray), weaker (light gray), and weak or absent (white) (more details linking the spectral broadening observed to the peak positions of the two modes are in SM.) (e) Full-width at half maximum ($\gamma$) of the peak at the expected UP2 anticrossing (magenta markers) as a function of the injected electrical power $P$. The dashed line is a fit using the formula given in Eq. 1, which accounts for Schawlow–Townes polariton narrowing and power-induced broadening to lowest order in $P$.

Polariton stimulation becomes relevant when the average number of polaritons per mode is comparable to, or exceeds, one. Estimating this requires the number of active modes, obtained from the density of states times the emission linewidth (SM). For our geometry this gives about 1500 polariton modes. We can then bound the occupation of the most populated mode. For a uniform distribution over all relevant modes and an emitted power of 10 nW (SM), the total number of polaritons is about 30, giving a lower bound $\sim 2 \times 10^{-2}$ per mode at an injected power of 0.15 W (electric field 300 V/cm). Conversely, if all polaritons occupy a single mode, the occupation reaches 0.3–1.3 at 1.45 W (electric field 1000 V/cm), depending on the cavity mode (inset of Fig. 3b and SM). This upper bound indicates proximity to threshold near 1.45 W, while the actual occupation lies between these limits.

To refine this estimate, we studied the dependence of cyclotron emission on injected electrical power (Fig. 3a). The calculated anticrossing curves assume an infinitely sharp CR, whereas experimentally (for more details see SM) its finite linewidth yields a broad profile whose high-energy tail activates the higher-energy emission feature slightly away from the ideal anticrossing (Fig. 3d). This explains the persistence of emission at 0.9 T, just outside the idealized anticrossing region of the second mode (SM). The emitted power in sample B grows superlinearly with injected electrical power from 0.15 to 1.45 W (Fig. 3b), suggesting that although lasing is not reached, the system may approach a stimulation threshold. The same nonlinear electroluminescence is reproduced in another cavity sample (sample D), but is absent in a control device without cavity (Fig. 3c). This shows that the effect is intrinsic to operation in the strong-coupling regime. A complementary estimate comes from the linewidth evolution (FWHM) versus injection. Figure 3e shows that emission in the spectral region of the second upper polariton branch in sample B narrows as $P$ increases, reaching a minimum near 0.5 W; for $P \gtrsim 0.8$ W it broadens above the low-power value. Such broadening may arise from nonlinear polariton losses [54], where a polariton converts into higher-energy Landau-level carrier excitations, as discussed above. We model the linewidth by

$$\gamma(P) = \frac{\gamma(0)}{1 + N_{pol}(P)} + \gamma_{NL}(P)\,, \tag{1}$$

where $N_{pol}(P)$ is the polariton occupation and $\gamma_{NL}(P)$ is an injection-induced broadening. The first term captures Schawlow–Townes narrowing consistent with possible stimulated scattering from final-state population, while the second accounts for nonlinear losses. To lowest order, $N_{pol}(P) \simeq \mathcal{N}P$ and $\gamma_{NL}(P) \simeq \eta P$, since both vanish as $P \to 0$. This expression fits the measured linewidth well (Fig. 3e) with $\gamma(0) = 5.65$ meV, $\mathcal{N} = 2.0$ W$^{-1}$ and $\eta = 3.1$ meV/W. At $P = 0.5$ W, we obtain $N_{pol} \simeq 1$, consistent with the bounds above.

Dirac–Landau polariton lasing has key advantages over bare cyclotron excitations. The polariton threshold is set by polariton losses, which can be reduced by improving both the cavity quality factor and cyclotron losses. At present the cavity quality factor (Q) is modest, $Q \approx 4$ (from reflectivity). Optimized THz structures with distributed Bragg mirrors, e.g. Tamm cavities [55], could greatly improve $Q$. Recent THz Tamm cavities reached $Q = 37$ at 1.5 THz [56], and Landau-polariton devices with distributed Bragg reflectors exceeded 500 [29]. Such improvements should lower the stimulation threshold and enable lasing before nonlinear losses dominate. Beyond HgTe QWs, strong coupling can also make viable materials that are otherwise unsuitable for streaming-effect stimulated emission [57] because they require prohibitively large electric fields. More broadly, many Dirac materials with non-equidistant Landau levels become realistic candidates for cyclotron-based light sources and lasers. In this context, spontaneous cyclotron emission in electrically biased monolayer graphene was recently observed in the mid-infrared [58], highlighting the potential for future THz and infrared emitters.

*Conclusions* — We demonstrated THz electroluminescence from Dirac–Landau polaritons in a strong-coupling regime between Landau-quantized Dirac fermions and cavity photons. This establishes a THz cavity-QED platform with relativistic electronic states. Emission occurs in a strongly non-equilibrium, nonlinear regime and shows spectral/intensity signatures consistent with possible stimulated polariton scattering, indicating proximity to polariton-lasing threshold. While lasing is not yet achieved, the results show a clear route to THz polariton lasers based on Dirac materials via improved cavity quality factors. The Dirac Landau-level structure reshapes polariton relaxation and scattering relative to conventional semiconductors and opens new directions to explore. This approach can be extended from HgTe QWs to a broader class of quantum materials hosting relativistic carriers.

This work was supported by the Terahertz Occitanie Platform, the French Agence Nationale pour la Recherche (TEASER, ANR-24-CE24-4830), the France 2030 program via Equipex+ HYBAT (ANR-21-ESRE-0026), the CNRS (Tremplin 2024 STEP), and the Russian Science Foundation (Project 24-62-00010). We thank E. Chauveau and A. Meguekam for help with substrate thinning, and J. Mangeney, T. Guillet, and J. Faist for discussions. We also thank B. Mongellaz, P. Lefebvre, and I. Philip for support with helium recovery. Finally, FT thanks C. Gimat for longstanding support.

# Supplementary Material for the article: "Nonlinear Terahertz electroluminescence from Dirac-Landau polaritons"

B. Benhamou-Bui[1], C. Consejo[1], S.S. Krishtopenko[1], S. Ruffenach[1], C. Bray[1], J. Torres[1], J. Dzian[2], F. Le Mardelé[2], A. Pagot[3], X. Baudry[3], S.V. Morozov[4,5,6], N.N. Mikhailov[7,8], S.A. Dvoretskii[8,9], B. Jouault[1], P. Ballet[3], M. Orlita[2], C. Ciuti[10], F. Teppe[1]

[1] *Laboratoire Charles Coulomb (L2C), UMR 5221 CNRS – Université de Montpellier, F-34095 Montpellier, France*
[2] *Laboratoire National des Champs Magnétiques Intenses, CNRS – UGA – UPS – INSA – EMFL, Grenoble, France*
[3] *CEA, LETI, MINATEC Campus, DOPT, Grenoble, France*
[4] *Institute for Physics of Microstructures of the Russian Academy of Sciences, Nizhny Novgorod, Russia*
[5] *Lobachevsky State University of Nizhny Novgorod, Nizhny Novgorod, Russia*
[6] *Ioffe Institute, 194021 St. Petersburg, Russia*
[7] *A.V. Rzhanov Institute of Semiconductor Physics, Siberian Branch of the Russian Academy of Sciences, Novosibirsk, Russia*
[8] *Novosibirsk State University, Novosibirsk, Russia*
[9] *Tomsk State University, Tomsk, Russia and*
[10] *Université Paris Cité, CNRS, Matériaux et Phénomènes Quantiques, 75013 Paris, France*

This Supplementary Material provides additional information on the experimental techniques, sample structure, cavity and polariton modeling, and data analysis methods that support the main findings of the manuscript. It includes: (i) a detailed description of the samples and the measurement protocols for electroluminescence and reflectivity; (ii) modeling and extraction procedures for Landau polariton coupling; (iii) quantitative estimates of the polariton population; and (iv) additional experimental data on the gate-voltage dependence of the emission and the cavity-induced spectral shaping.

## METHODS

### Samples

Two types of samples were studied, both grown via Molecular Beam Epitaxy (MBE) on GaAs-(013) or CdTe-(100) substrates (see Table I for details). Two distinct buffer layers were employed, as illustrated in Figure S.1. The active layer consists of a HgTe QW with HgCdTe barriers, with only one sample type incorporating a CdTe cap layer. To create an optical cavity, the substrates were thinned below 50 $\mu$m using two techniques: (i) micrometric rotary sawing down to 40 $\mu$m or (ii) mechanical polishing to 50 $\mu$m, followed by inductively coupled plasma etching to further reduce the thickness below 30 $\mu$m. The samples were then mounted on a gold sample holder.

### Landau emission measurement technique

Higher Landau levels are populated using short electrical pulses with a frequency of 127 Hz, peak-to-peak amplitudes ranging from 40 V to 200 V, and durations between 1 ms and 30 $\mu$s. These pulses are applied to the sample via indium balls soldered onto its surface, which diffuse into the structure, forming high-quality ohmic contacts. The experimental setup includes a Landau spectrometer with three superconducting coils housed in a liquid helium cryostat. The detector is an $n$-type InSb bolometer, which operates under a strong magnetic field to refine and narrow its detection energy window. The detector has a responsivity of $10^4$ V/W. The second coil resolves the LL structure of the sample, while the third coil decouples the contributions of the first two fields.

The measurement protocol involves fixing the sample's magnetic field while sweeping the detector's magnetic field, enabling energy spectra acquisition at a constant sample field. The detector signal is amplified by a low-noise amplifier and processed via a boxcar averager, ensuring a high signal-to-noise ratio even for low-duty-cycle pulses ranging from 0.4% to 1%. All measurements were performed at 4.2 K.

### Cavity characterization

When the substrate is thin, i.e. on the order of $\lambda/2$ for the cyclotron wavelength, the emission spectrum is strongly modified by the presence of cavity electromagnetic modes. The metal contact placed on the back side of the sample and the helium/semiconductor interface on the front side act as mirrors for a vertical resonant THz cavity. In order to characterize the cavity effect, we performed emission measurements at zero magnetic field. When the LLs are not yet established, the emission spectrum is not related with the CR and is primarily composed of blackbody radiation, most likely originating from the heating of the current injection contacts. When the dimensions of a thermally emitting object are on the order of the thermal radiation wavelength $\lambda_{Th}$, its emission can indeed substantially differ from the predictions made by Planck's law [1, 2], creating new possibilities in the realm of thermal radiation. For instance, the presence of some Fabry-Pérot cavity modes can greatly enhance thermal emission, producing a narrow-band emission spectrum consistent with the Purcell effect. Moreover, the spectral emission power can exceed the limits imposed by Planck's law for blackbody radiation [3].

| Sample name | Substrate | Thickness | Cavity mode energies | Carriers density [cm$^{-2}$] |
|---|---|---|---|---|
| Sample A | GaAs | 28 $\mu$m | 3.1 meV ; 10.6 meV | $7 \times 10^{11}$ |
| Sample B | GaAs | 30 $\mu$m | 3.1 meV ; 10.1 meV | $4 \times 10^{11}$ |
| Sample C | CdTe | 40 $\mu$m | 5 meV ; 9 meV | $3 \times 10^{11}$ |
| Sample D | GaAs | 28 $\mu$m | 3.1 meV ; 10.6 meV | $3 \times 10^{11}$ |

TABLE I. Table summarizing the different samples' characteristics.

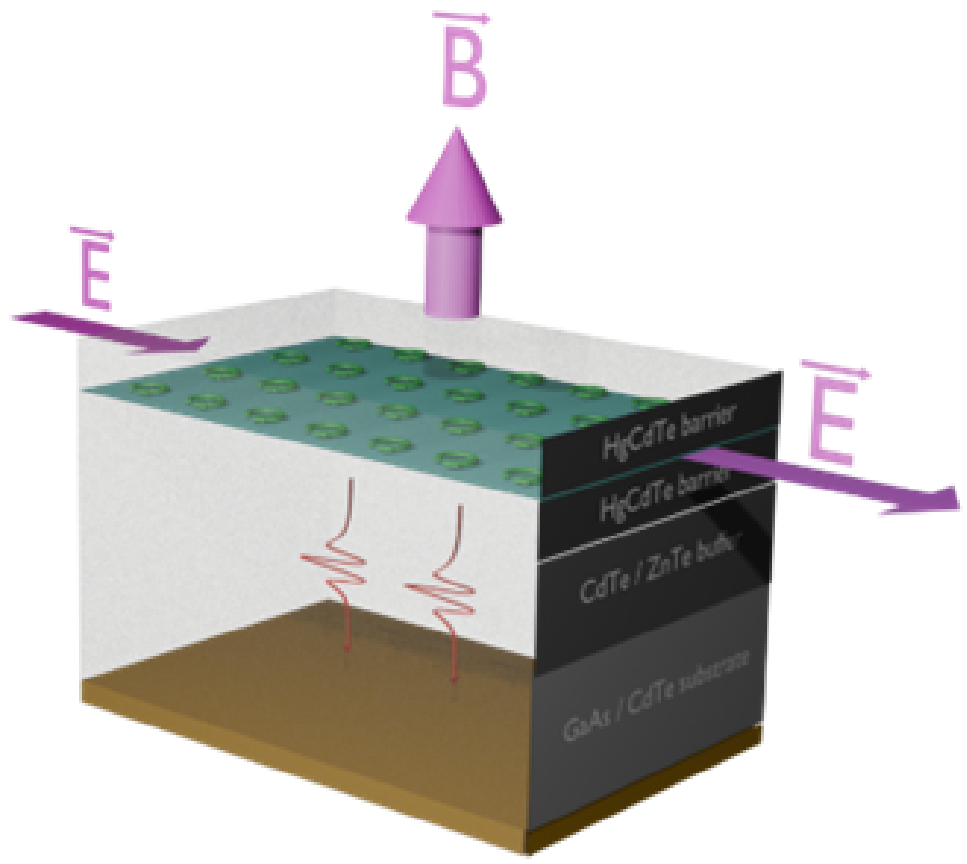

FIG. S.1. **Schematic representation of the sample layer structure.** The core consists of a HgTe quantum well (QW) sandwiched between HgCdTe barriers with a Cd composition of 68%. A CdTe/ZnTe buffer layer is inserted beneath the QW to facilitate strain relaxation before reaching the substrate.

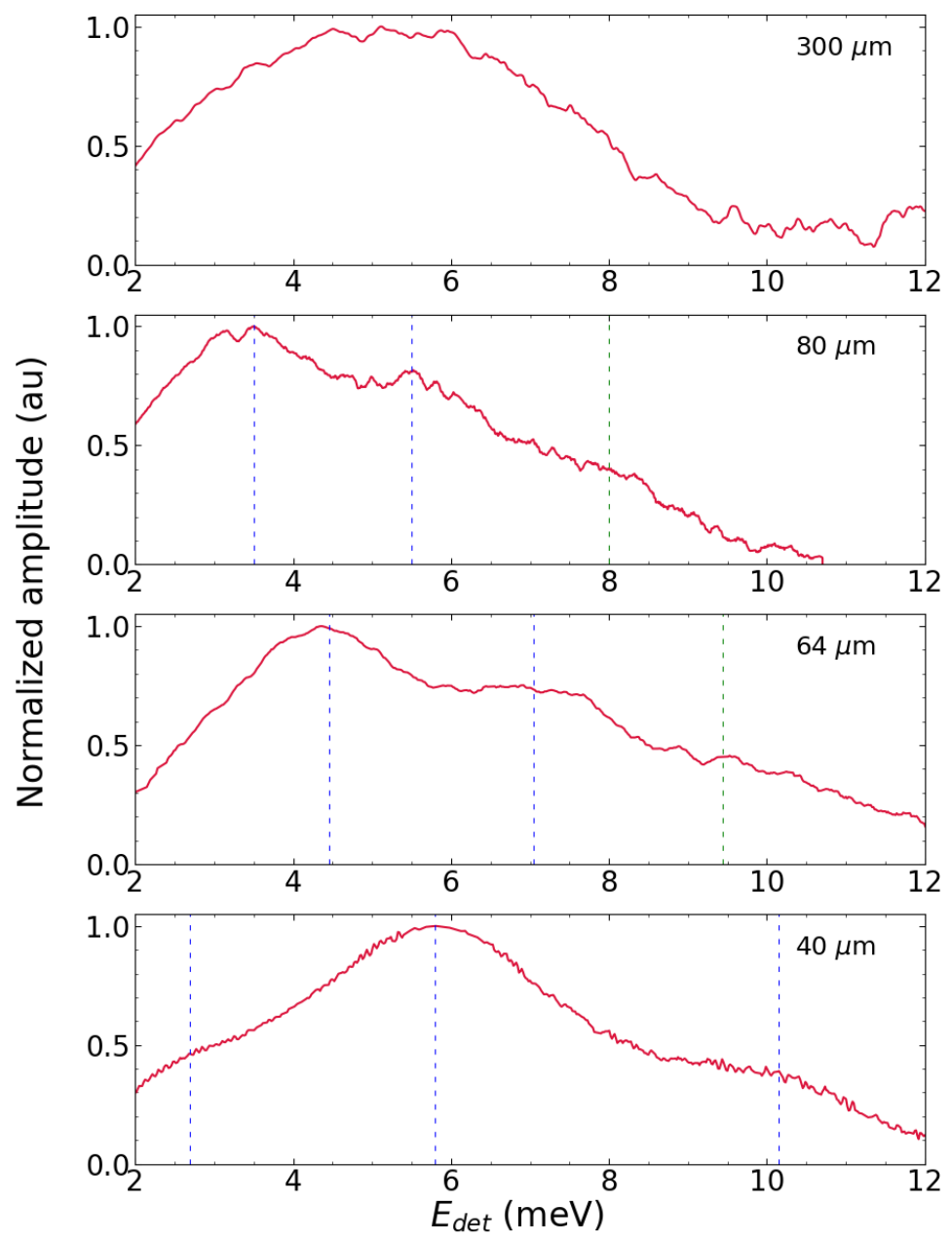

FIG. S.2. **Emission spectra obtained at** 4.2 **K, at zero magnetic field, on sample A, for different substrate thicknesses.** On the bare sample (300 $\mu$m), the spectrum exhibits no structuration. When the substrate thickness is reduced below 100 $\mu$m, additional extrema (blue dashed lines) emerge, indicating a cavity effect. Notably, the 8 meV line in the 80 $\mu$m sample and the 9.5 meV line in the 64 $\mu$m sample (grey dashed lines) are barely visible at $B = 0$ T but become observable through cyclotron resonance at finite magnetic fields.

Figure S.2 below shows these emission spectra plotted for the different sample thicknesses. When the substrate is thick, the blackbody emission spectrum is broadband, and its peak is, as expected, red shifted as the injection current decreases. When the substrate is reduced to thicknesses comparable to the wavelength of thermal radiation defined by Wien's displacement law, the broadband blackbody radiation transforms, uncovering peaks at energies aligned with the substrate's optical cavity modes. The interaction between the resonant modes of the cavity induces a narrower thermal emission bandwidth [4, 5]. Indeed, since the sample acts at zero magnetic field as both a thermal heat source and an optical cavity, its thermal radiation is modulated by the cavity modes according to the Purcell effect.

The emission signal is structured into a series of peaks, with their positions varying depending on the substrate thickness. We then compare in Figure S.3 these energy positions with a simulation using the Transfer Matrix Method (TMM) [6]. Taking into account the sample geometry (thinned substrate on a gold sample holder) and the complex refractive index of the material, we were able to accurately reconstruct the energy positions of the modes for the different thicknesses. There is excellent agreement between the experimental data and the simulation, as the theoretical thicknesses used to reproduce the results closely match those measured experimentally.

Another way to characterize the cavity effect induced by the substrate is to perform room temperature THz reflectivity thanks to a commercial Time-Domain-Spectroscopy (TDS) setup. Figure S.4 below shows two spectra obtained on the two thinnest samples. It clearly displays some reflectivity minima, well corresponding to the cavity modes. These minima positions, when extracted, are once again in a very good agreement with the TMM simulations.

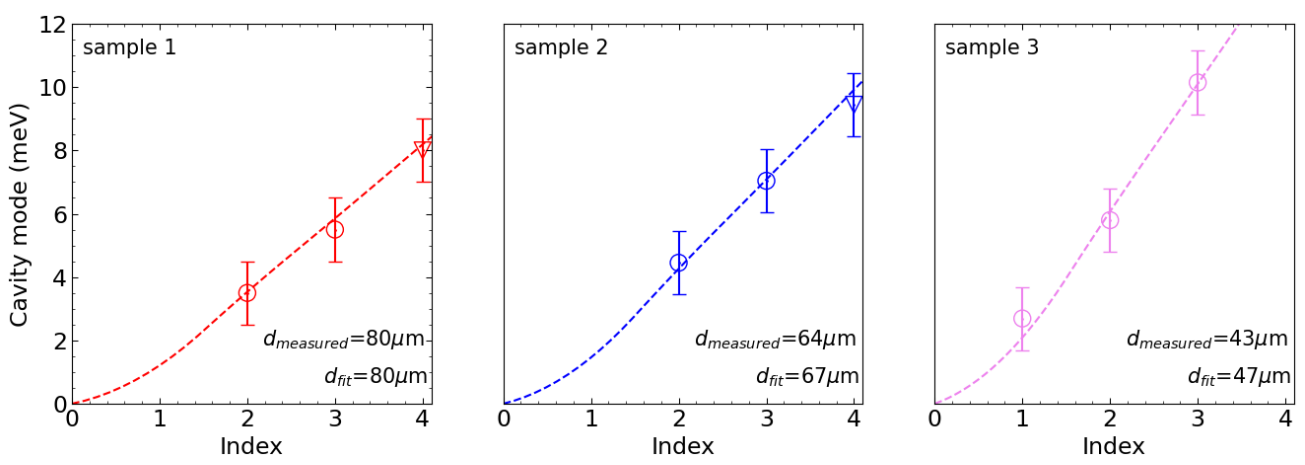

FIG. S.3. **Evolution of cavity mode energies extracted from Fig. SM2 for three different samples with varying substrate thicknesses (empty circles).** Triangular markers correspond to data obtained at high magnetic fields. The energy values were deduced from the $B = 0$ T emission spectra. A Transfer Matrix Method (TMM) prediction (dashed lines) is overlaid, showing excellent agreement with the experimental data.

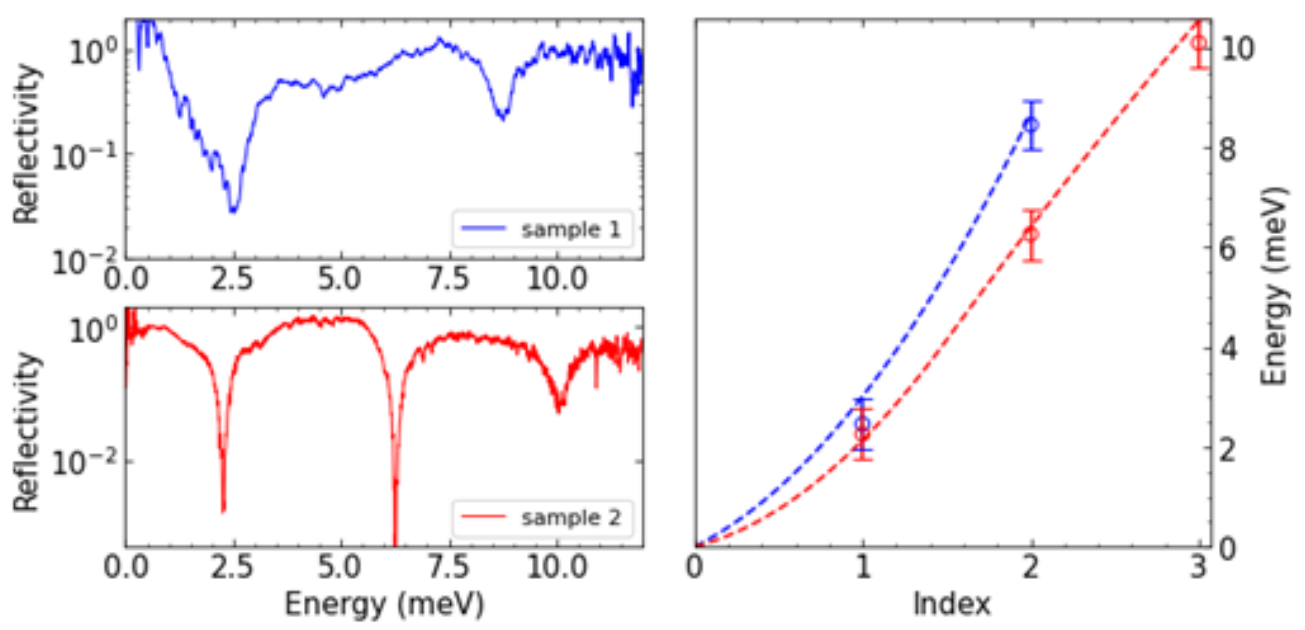

FIG. S.4. **THz reflectivity and transmission analysis.** (Left panel) TDS reflectivity spectra measured for samples A and B. (Right panel) Transmission minima extracted and plotted against an arbitrary index (empty circles). A TMM-based prediction (dashed lines) is overlaid, showing excellent agreement with the experimental data.

## Magneto-transport measurements – Shubnikov-de Haas regime

The electron densities of the samples are determined using the standard Shubnikov-de Haas magneto-transport technique. Figure S.5 below presents the typical magneto-resistance behavior of our samples. The observed oscillations enable us to extract the electron density via the well-known formula:

$$n_S = \frac{e}{h \cdot \Delta(1/B)}\,, \tag{S.1}$$

where $\Delta(1/B)$ is the inverse magnetic field period of the oscillations. For the sample shown below, this yields $n_S = 7.0 \times 10^{11}$ cm$^{-2}$.

Additionally, cyclotron resonance measurements allow us to determine the cyclotron mass of the sample (see main text). By combining these results, we can verify the predicted evolution of the cyclotron mass as a function of electron density, as derived from the low-energy model[7]. The inset demonstrates excellent agreement between the experimental data and the theoretical model.

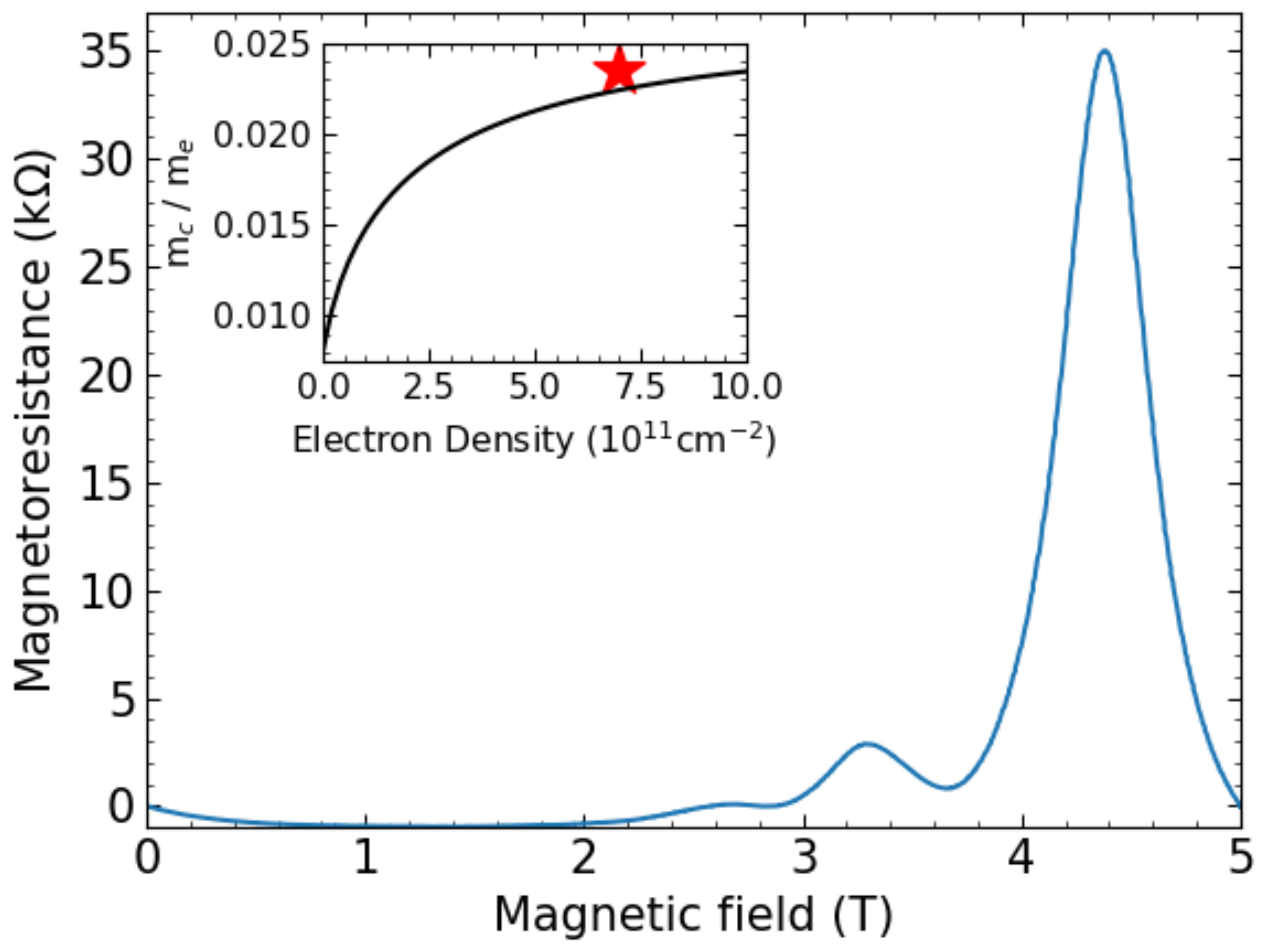

FIG. S.5. **Longitudinal magneto-resistance and cyclotron mass evolution.** Shubnikov-de Haas oscillations measured in the longitudinal magneto-resistance of sample A. (Inset) Theoretical dependence of the cyclotron mass on electron density, based on a low-energy model [7]. The red star represents the experimental value extracted from reflectivity and magneto-transport measurements, showing consistency with theoretical predictions.

Regarding the magneto-reflectivity and cyclotron emission results, it is important to note that the system operates within the Shubnikov-de Haas regime. This implies that the Landau levels (LLs) are already present but still exhibit partial overlap. These conditions, which lie between the classical and quantum regimes, are often referred to as the incipient Landau quantization regime.

Similar to the classical case, the cyclotron resonance (CR) energy evolves linearly with the applied magnetic field. This occurs because, as the magnetic field increases, the Fermi level oscillates between adjacent LLs. Consequently, even though the LLs in the system are relativistic and follow a $\sqrt{B}$ dependence, the energy of optical transitions between these levels remains linear in the magnetic field. This can be understood by considering the energy difference between two adjacent LLs, $n+1$ and $n$:

$$\hbar\omega_c = \sqrt{2v_F^2 eB\,\hbar}\,\left(\sqrt{n+1} - \sqrt{n}\right) \text{ with } E_F = \sqrt{2v_F^2 eB\,\hbar}\sqrt{n}\,. \tag{S.2}$$

Considering that for $n \gg 1$

$$\sqrt{n+1} - \sqrt{n} \simeq \frac{1}{2\sqrt{n}}\,, \tag{S.3}$$

we finally have

$$\begin{aligned} \hbar\omega_c &= \sqrt{2v_F^2 eB\,\hbar}\,\left(\sqrt{n+1} - \sqrt{n}\right) \\ &\simeq \sqrt{2v_F^2 eB\,\hbar}\frac{1}{2\sqrt{n}} \\ &= \frac{2v_F^2 eB\,\hbar}{2E_F} = \frac{eB\,\hbar}{m_c}\,, \end{aligned} \tag{S.4}$$

where we defined $m_c = \frac{E_F}{v_F^2}$.

**Landau polariton: Fitting procedure**

To extract the coupling strength from the reflectivity measurements, we followed the same procedure as in [8]. It is based on the following total Hopfield Hamiltonian [9]:

$$H = H_{cavity} + H_{Landau} + H_{int} + H_{dia}\,, \tag{S.5}$$

where $H_{cavity}$ is the bare Hamiltonian of the cavity, $H_{Landau}$ describes the collective cyclotron excitation of the electrons occupying the Landau levels, $H_{int}$ is the paramagnetic light-matter interaction, while $H_{dia}$ is the diamagnetic contribution. The polariton excitations can be obtained by diagonalizing the following Hopfield-Bogoliubov matrix:

$$M_j(B,\chi_j) = \hbar \begin{pmatrix} \omega_c & \chi_j\sqrt{\omega_c} & 0 & \chi_j\sqrt{\omega_c} \\ \chi_j\sqrt{\omega_c} & \omega_j + 2\chi_j^2 & \chi_j\sqrt{\omega_c} & 2\chi_j^2 \\ 0 & -\chi_j\sqrt{\omega_c} & -\omega_c & -\chi_j\sqrt{\omega_c} \\ -\chi_j\sqrt{\omega_c} & -2\chi_j^2 & -\chi_j\sqrt{\omega_c} & -\omega_j - 2\chi_j^2 \end{pmatrix}\,, \tag{S.6}$$

where

$$\chi_j = \frac{\Omega_j}{\sqrt{\omega_c}} \tag{S.7}$$

is a fitting parameter, independent of the magnetic field, $\Omega_j$ being the collective polariton coupling (Rabi) frequency for the electromagnetic mode $j$. By diagonalizing this Hamiltonian, we can access the theoretical polaritonic branches, labelled $\omega_j^{UP}(B,\chi)$ (resp. $\omega_j^{LP}(B,\chi)$ ) for the upper branch (resp. lower branch). Therefore, we can extract the coupling strength by minimizing the quantity

$$RMSD_j(\chi) = \sqrt{\frac{\sum_{\chi}^{N_{exp}}\left[\left(\omega_{j,\eta}^{UP} - \omega_j^{UP}(B_\eta,\chi)\right)^2 + \left(\omega_{j,\eta}^{LP} - \omega_j^{LP}(B_\eta,\chi)\right)^2\right]}{2N_{exp}}}\,, \tag{S.8}$$

where $N_{exp}$ is the number of experimentally measured points. One example of minimization is shown in Fig. S.6.

**Estimation of the number of polaritons per mode**

*Lower bound estimate*

Let us consider the cavity energy dispersion

$$E_{cav}(k) = \frac{\hbar c}{n}\sqrt{k_z^2 + k^2}\,, \tag{S.9}$$

where $c$ is the speed of light, $n$ the refractive index of the substrate, z is the direction perpendicular to the 2D material. For small in-plane wave vectors, the cavity dispersion can be approximated as parabolic, namely:

$$E_{cav}(k) = \frac{\hbar c}{n}k_z\sqrt{1 + \frac{k^2}{k_z^2}} \simeq E_{cav}(0)\left(1 + \frac{1}{2}\frac{k^2}{k_z^2}\right)\,. \tag{S.10}$$

The density of photonic states is

$$\begin{aligned} \mathcal{D}(E) &\simeq \frac{A}{(2\pi)^2}\int d^2k\,\delta(E - E_{cav}(k)) \\ &= \frac{A}{2\pi}\int kdk\,\delta(E - E_{cav}(k))\,, \end{aligned} \tag{S.11}$$

where A is the area of sample where there is emission. Since $dE_{cav} = E_{cav}(0)\frac{kdk}{k_z^2}$, then $kdk = \frac{k_z^2}{E_{cav}(0)}dE_{cav}$ and therefore

$$\mathcal{D}(E) \simeq \frac{A}{2\pi}\frac{k_z^2}{E_{cav}(0)}\,. \tag{S.12}$$

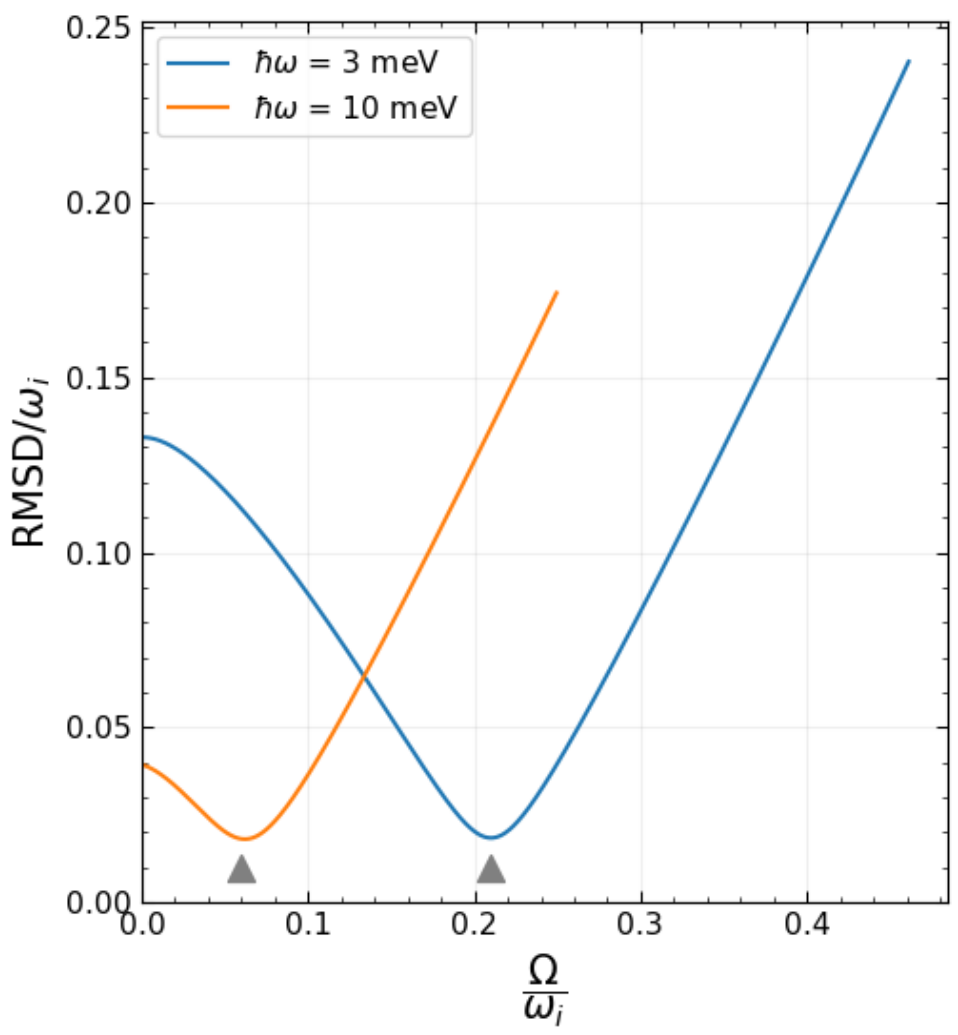

FIG. S.6. **Normalized root-mean-square deviation analysis for sample A.** Evolution of the normalized root-mean-square deviation for the two electromagnetic modes. The minima correspond to normalized coupling strengths of 21% and 4% for the lower and higher modes, respectively, indicating distinct interaction regimes.

The polariton density of states is comparable to the photonic density of states, which will be used for our estimate of the number of polariton modes involved in the emission process:

$$\mathcal{N}_{mode} \simeq \mathcal{D}(E)\,\Delta E_{emission} = \frac{A}{2\pi} k_z^2 \frac{\Delta E_{emission}}{E_{cav}(0)}\,, \quad \text{(S.13)}$$

where $\Delta E_{em}$ is the emission linewidth. Given the experimental nominal parameters, we can take $A \simeq 1\,mm^2$, $k_z = \frac{\pi}{L_{cav}} = \frac{\pi}{30\mu m} = 10^5\,m^{-1}$ and $\Delta E_{emission} = 3\,\text{meV}$ we have $E_{cav}(0) \simeq 3\,\text{meV}$ we get $\mathcal{N}_{mode} \simeq 1500$. Finally, we can estimate the number of polariton via the emitted power:

$$\mathcal{P}^{emission} \simeq N_{pol}\,\gamma_{rad}^{avg}\,\hbar\omega_{pola}^{avg}\,, \quad \text{(S.14)}$$

where $N_{pol}$ is the number of polaritons in the steady state, $\gamma_{rad}^{avg}$ the average radiative rate and $\hbar\omega_{pola}^{avg}$ the average photon emission energy. By injecting the experimental values $\mathcal{P}^{emission} \simeq 10\,nW$, $\hbar\omega_{pola}^{avg} \simeq 3\,\text{meV}$ and $\gamma_{rad}^{avg} \simeq \frac{\Delta E_{emission}}{h} = \frac{3\,\text{meV}}{h} \simeq 0.7\,ps^{-1}$ we get: $N_{pol} \simeq 30$. Finally, we get the lower bound:

$$\frac{N_{pol}}{\mathcal{N}_{mode}} > 2 \cdot 10^{-2}\,. \quad \text{(S.15)}$$

*Upper bound estimate*

The upper bound for the polariton occupation number is obtained by assuming that all the polaritons are occupying the same polariton mode. This is certainly not the case, but together with the lower bound calculated above will allow us to have a decent estimate of the polariton occupation numbers. Assuming that only one mode participates to the emission, we can estimate the polariton population from the non-linear dependence of the emission amplitude with respect to the injected electrical power. This can be calculated via the simple rate equation:

$$\frac{dN_{pol}}{dt} \simeq -\gamma N_{pol} + \eta P_{elec}(1 + N_{pol})\,, \quad \text{(S.16)}$$

where $\gamma$ is the polariton loss rate and $\eta$ is unknown. The steady-state solution reads:

$$N_{pol} = \frac{\eta P_{elec}}{\gamma - \eta P_{elec}}\,. \quad \text{(S.17)}$$

From this equation we can see that for $P_{elec} \to P_{elec}^{(thresholds)} = \frac{\gamma}{\eta}$ then $N_{pol} \to +\infty$. Below threshold, we can Taylor-expand the previous solution as follows:

$$N_{pol} = \frac{\eta P_{elec}}{\gamma} \frac{1}{1 - \frac{\eta P_{elec}}{\gamma}} \simeq \frac{\eta P_{elec}}{\gamma}\left(1 + \frac{\eta P_{elec}}{\gamma}\right) \simeq \frac{\eta P_{elec}}{\gamma} + \left(\frac{\eta P_{elec}}{\gamma}\right)^2 \quad \text{(S.18)}$$

Therefore, we now have the dependence of the emission power on the electrical power, namely

$$\mathcal{P}^{emission} \simeq N_{pol}\,\gamma_{rad}^{avg}\,\hbar\omega_{pola}^{avg} \simeq \gamma_{rad}^{avg}\,\hbar\omega_{pola}^{avg}\left[\frac{\eta P_{elec}}{\gamma} + \left(\frac{\eta P_{elec}}{\gamma}\right)^2\right] = a_1 P_{elec} + a_2 P_{elec}^2\,, \quad \text{(S.19)}$$

where $a_2/a_1 = \eta/\gamma$. We can therefore have access to the polariton population by fitting the curve corresponding to the measured emitted power versus the electrical injected power, as shown in the figure 3 in the main text. For the considered sample, the polariton population would reach more than one for an electrical injected power of 1.5 W (obtained for an electric field of 1 kV/cm).

From the two scenarios developed above, we can conclude that the polariton occupation from the most populated mode in our system is bound as follows:

$$2 \cdot 10^{-2} \leq N_{pol} \leq\simeq 1\,. \quad \text{(S.20)}$$

### Emission / transmission comparison

The figure S.7 shows an emission spectrum on top of reflectivity one, for the same sample and magnetic field value. It highlights the fact that the emission maximum is aligned with the upper branch of the Landau polariton and that the lower branch remains invisible in emission.

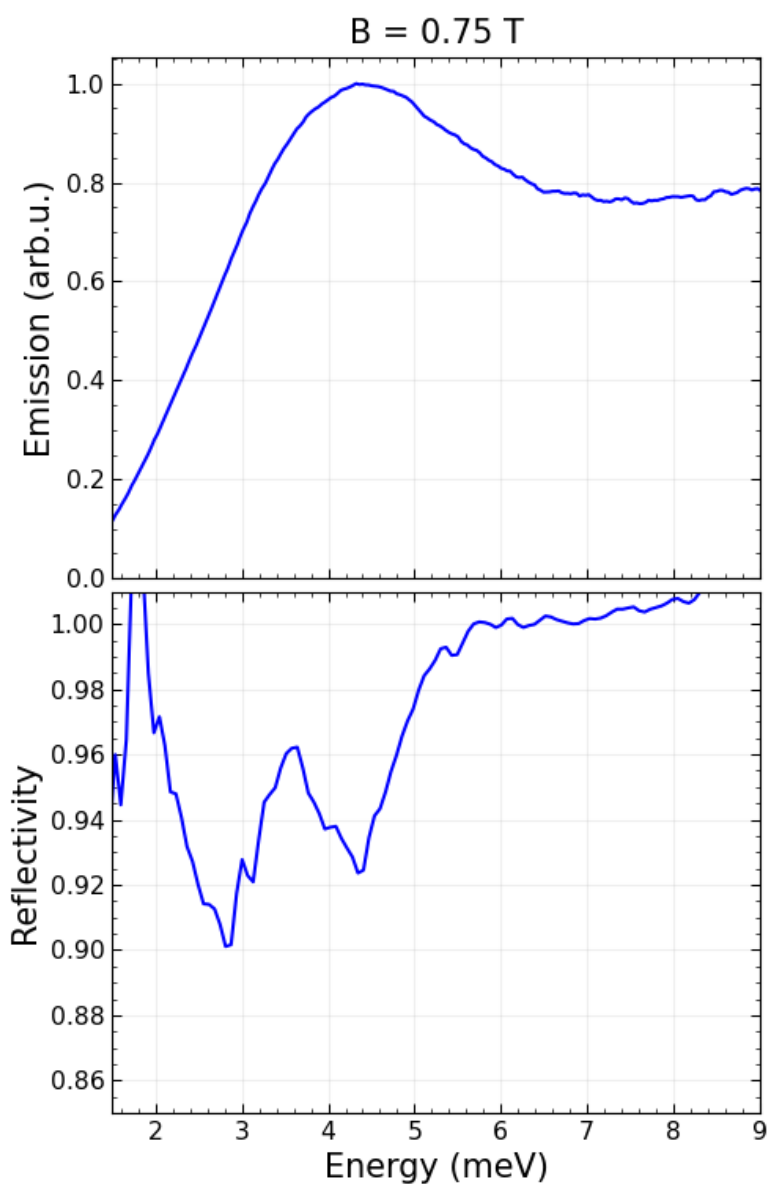

FIG. S.7. **Comparison of emission and transmission spectra at 0.75 T for sample A.** (Top) Emission spectrum measured at a magnetic field of 0.75 T. (Bottom) Corresponding transmission spectrum obtained under the same conditions. The data clearly reveal that only the upper branch of the Landau polariton exhibits significant emission, highlighting the asymmetric population of polariton branches.

Indeed, the higher polariton branch dominates in our emission spectra. This may result from a bottleneck effect that strongly inhibits the redistribution of polaritons from the upper to the lower branches. Such a bottleneck arises from a complex interplay of relaxation mechanisms, including polariton–phonon, polariton–carrier, and polariton–polariton scattering, as well as intrinsic radiative and non-radiative losses. While near-infrared exciton-polaritons typically relax efficiently toward the lower branch, our Landau-quantized system exhibits the opposite trend, with electroluminescence predominantly emitted from the upper branches. A possible explanation lies in the combined influence of non-equidistant Landau levels and the non-equilibrium carrier distribution generated by electrical injection, especially carriers occupying high-energy states above the Fermi level. These electrons can absorb energy at lower scales due to the reduced spacing between Landau level. This effect may be further amplified by the presence of crossed electric and magnetic fields, which cause higher energy levels to shift downward more rapidly than lower ones. In addition, Coulomb interactions may enhance this process by enabling the conversion of polaritons into electron–hole excitations at similar energies through interactions with non-equilibrium carriers.

## Gate voltage effect

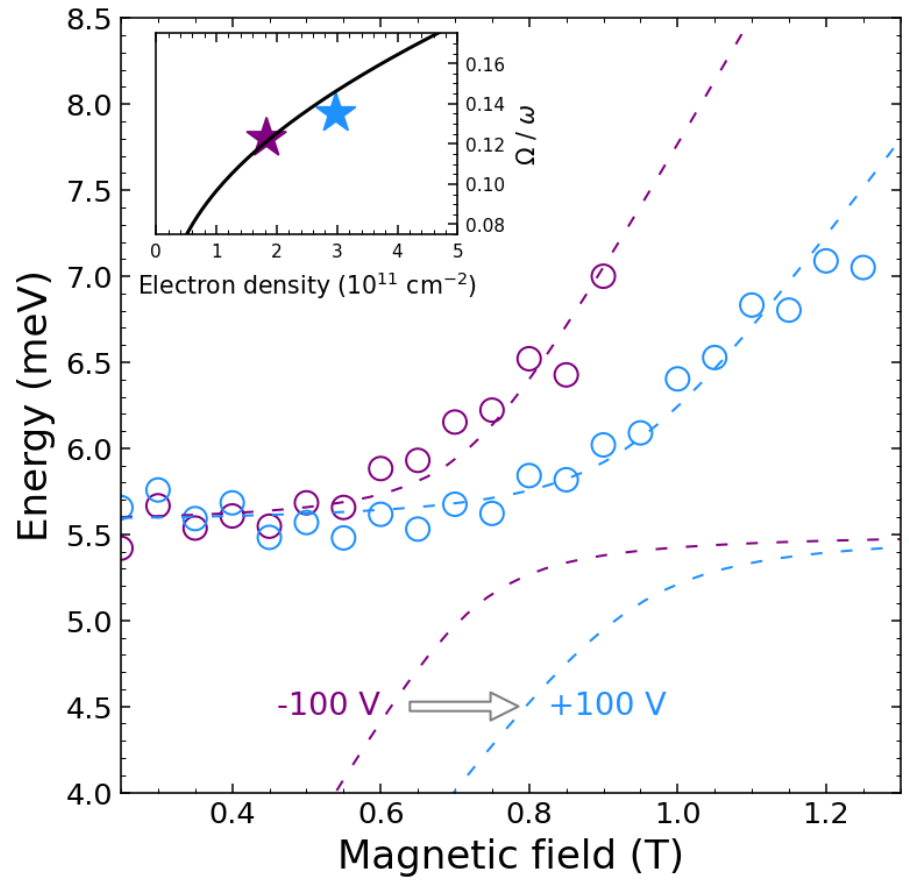

FIG. S.8. **Influence of the gate voltage on coupling strength and critical magnetic field.** Due to the relativistic nature of charge carriers, their concentration in the QW determines the cyclotron mass and thus the slope of the Landau emission. By tuning the carrier density from $n_s = 1.83 \times 10^{11}$ cm$^{-2}$ to $2.98 \times 10^{11}$ cm$^{-2}$, the cyclotron mass increases, shifting the critical magnetic field for the anticrossing from approximately 0.7 T to 0.9 T. (Main) Extracted emission maxima from spectra measured at different magnetic field values for sample C, with a gate bias of −100 V (purple empty circles) and +100 V (blue empty circles). The dashed lines represent fits based on the model developed in Ref.8. (Inset) Evolution of the reduced coupling strength as a function of electron density (black line), based on the model in Ref.8 and the low-energy model describing the cyclotron mass dependence on density [10]. The violet (resp. blue) star represents the extracted reduced coupling strength from the emission curve for a gate bias of −100 V (resp. +100 V). The model has been rescaled to match the experimental values.

Dirac-Landau polaritons exhibit distinctive features compared to the standard Landau polaritons. Because the effective mass of Dirac electrons depends on the electron density, the cyclotron mass and, consequently, the CR slope are continuously tunable via the back-gate voltage. Therefore, the critical magnetic field at which the anticrossing occurs can also be tuned by adjusting the gate voltage (see S.8). Even more interestingly, since the coupling strength term depends on the cyclotron mass, it can be tuned as well with the gate voltage, as illustrated in the insert of S.8. According to the Hopfield polariton model, where the cyclotron mass is derived from the Bernevig-Hughes-Zhang (BHZ) model, the coupling strength can reach 30% in this non-optimized system, for a realistic carrier density of $1 \times 10^{12}$ cm$^{-2}$.

The figure S.9 displays two raw emission spectra obtained for two extremes gate bias value of +100 and -100 V and for the same magnetic field value. It highlights the specificity of Dirac materials which is a density-dependent mass inducing a shift of the CR energy and therefore a shift of the Landau polariton anti-crossing.

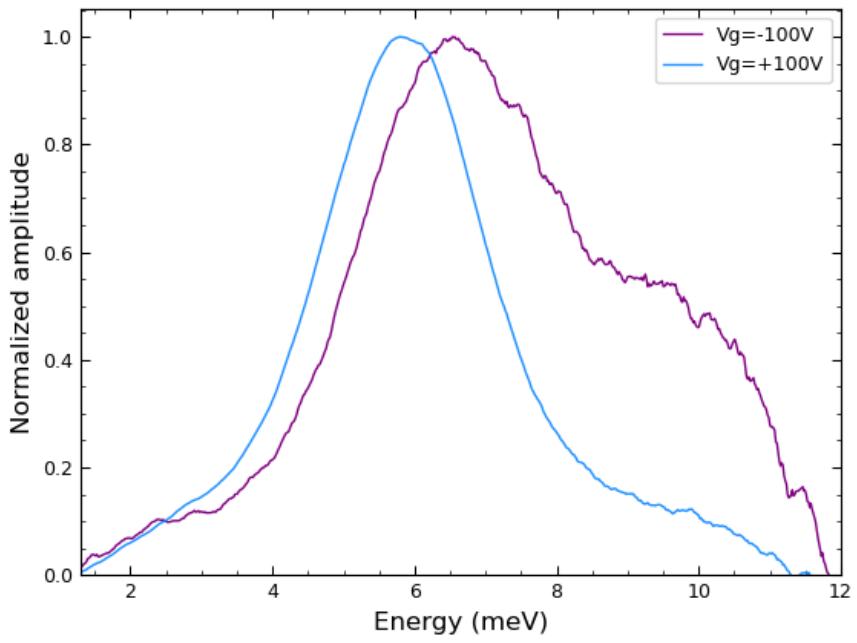

FIG. S.9. **Gate voltage influence on Landau polariton emission.** Emission spectra recorded at a magnetic field of 0.8 T for sample B, with a gate bias of −100 V (purple curve) and +100 V (blue curve). As the bias shifts from −100 V to +100 V, the upper branch of the Landau polariton undergoes a redshift of nearly 1 meV, attributed to the increase in cyclotron mass with higher electron density.

### Amplitude and FWHM extraction

To extract the amplitude and the FWHM of the different emission peaks, we used a double-Gaussian fit for every injection bias value. The figure S.10 shows the 13 fits obtained on the experimental spectra.

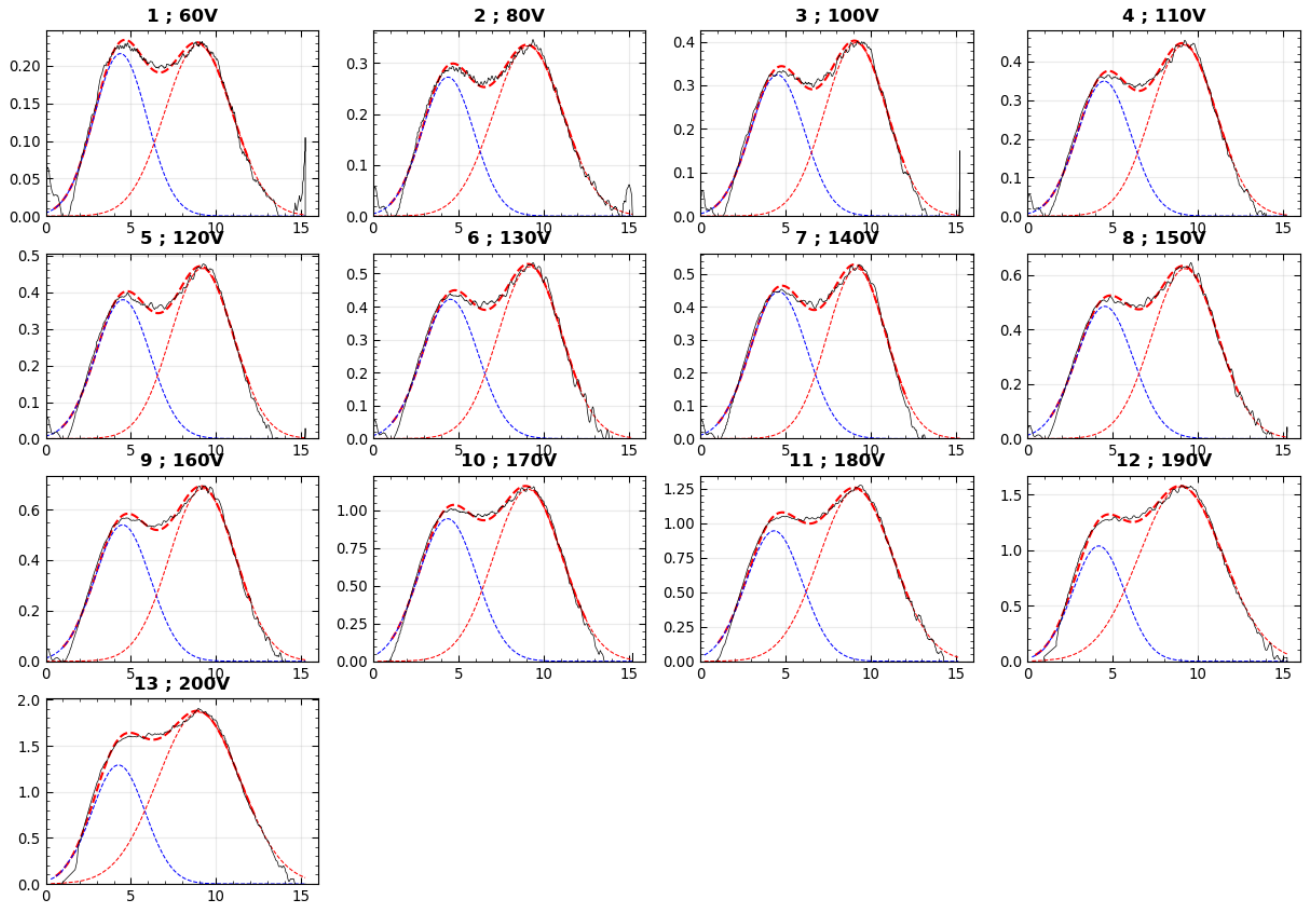

FIG. S.10. **Bias-dependent emission spectra at 0.9 T for sample A.** Emission spectra recorded for different injected bias values. The total double-Gaussian fit is overlaid (red dashed line), along with the individual Gaussian components corresponding to the UP1 (blue dotted line) and UP2 (red dotted line) polariton branches.

### Thermal background in cyclotron emission spectra

We note that the low-energy tail observed at high magnetic fields in the emission spectra originates from a residual thermal background signal. Although this contribution is relatively weak, it becomes more pronounced due to the normalization applied to all spectra for visual clarity. This normalization, scaling each spectrum to its maximum, tends to amplify small residual variations, especially at high magnetic fields where the cyclotron emission weakens. As discussed in Gebert et al. [7], this decrease arises from a combination of factors, including the proximity to the Reststrahlen band, limitations in the detector's spectral response, and, most notably, a non-equilibrium carrier distribution resulting from electric field excitation, which modifies the occupation of Landau levels relative to a Fermi–Dirac distribution at thermal equilibrium. At low magnetic fields, the emission peak appears asymmetric, with a pronounced high-energy tail. This is primarily due to the detector's poor sensitivity below 2 meV and the lack of correction for its spectral response, which distorts the true (symmetric) line shape. The effect is further enhanced by increased thermal background at low photon energies. This background, likely originating from Joule heating at the emitter contacts during excitation, was consistently observed across all samples. To minimize its impact, the signal measured outside the electrical excitation window was subtracted from the signal during excitation using a boxcar integrator. While this approach effectively suppresses most of the background, it does not fully eliminate it. The residual contribution is especially visible at low magnetic fields, where the lower sample resistance leads to higher current injection at a given voltage, enhancing contact heating and the associated thermal emission.

### I(V) characteristic of the contacts

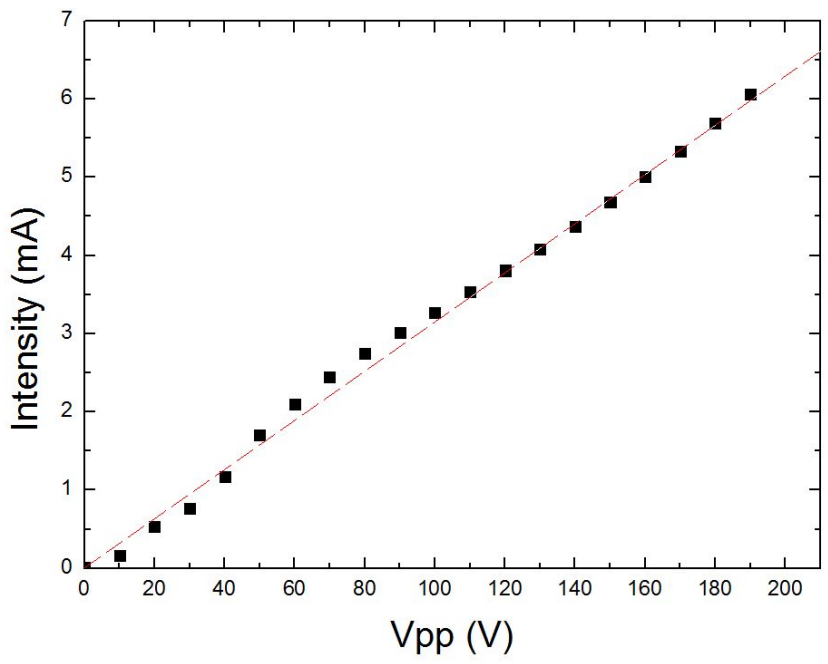

FIG. S.11. I(V) characteristic measured on sample A of the main text, showing an ohmic behavior (red dashed line).

To ensure that we obtain ohmic contacts when welding

Indium balls on our samples, we performed I(V) measurements. Figure S.11 shows an example of such a characteristic curve, displaying a good linear behavior. It was obtained by including a shunt resistor of 1 Ohm on one of the injection arm of our set-up. By measuring the voltage drop at its terminals, we can have access to the current flowing across the sample.

### Influence of the cavity on the emission's amplitude

The observed increase in electroluminescence amplitude with improved cavity confinement is consistent with the expected enhancement from optical resonators. However, quantitative analysis is limited by experimental uncertainties, including sample positioning, coupling to oversized waveguides, and variability in manually applied indium contacts. These factors affect injection efficiency and signal collection, making direct amplitude comparisons across samples unreliable.

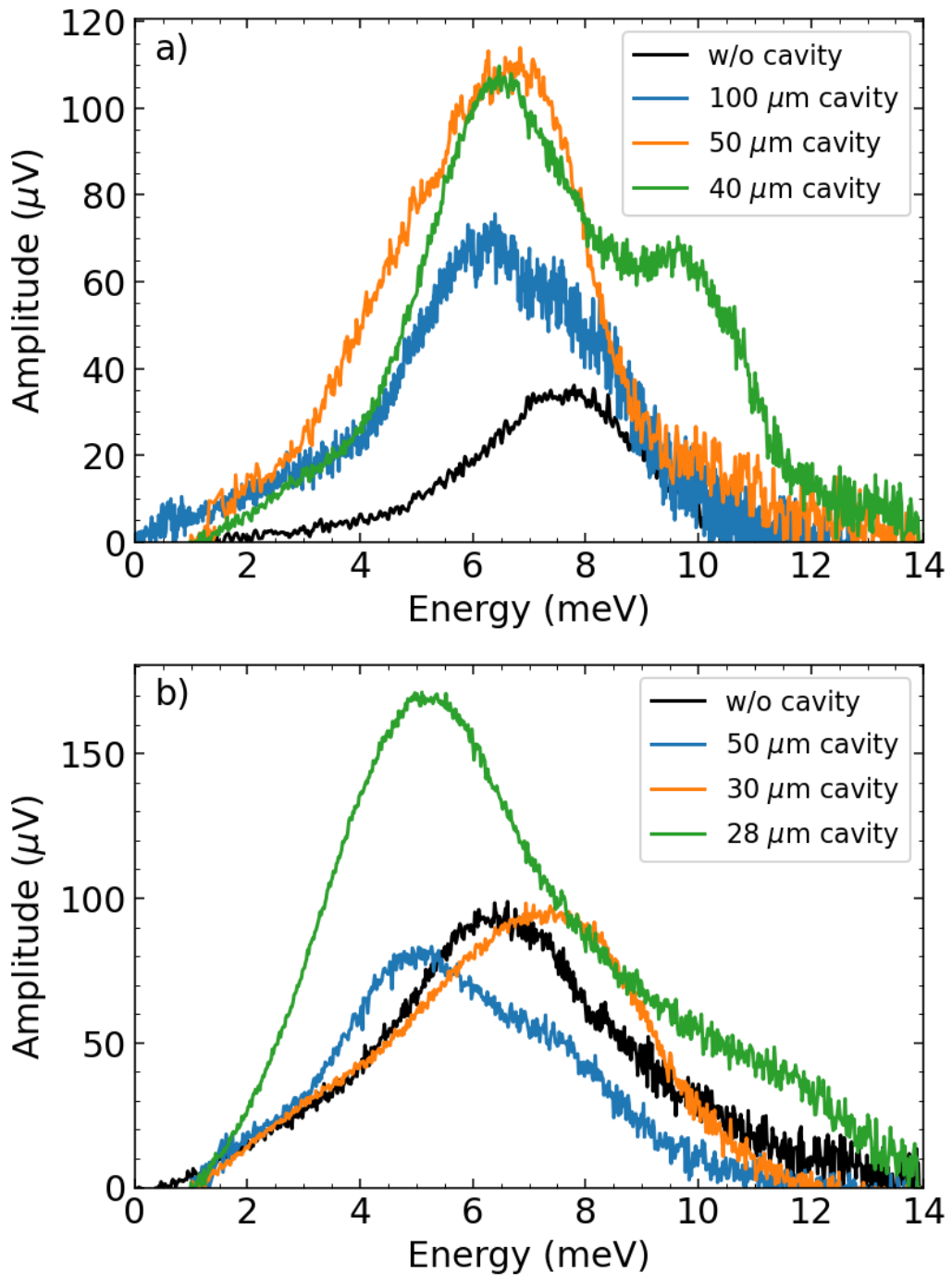

FIG. S.12. Electroluminescence spectra recorded from samples with varying cavity sizes under a magnetic field of 1 T (colored curves). The spectra shown in panel a (respectively, panel b) were measured from the growth corresponding to sample C (respectively, samples A and B) discussed in the main text. For comparison, each panel also includes a spectrum from a reference sample without a cavity (black curves).

### Influence of he injected electrical power

To make sure that the non-linear evolution of emitted power with the injected electrical power is indeed induced by the strong-coupling regime, we performed a similar experiment on a bare sample. The results are shown in Fig. S.13.

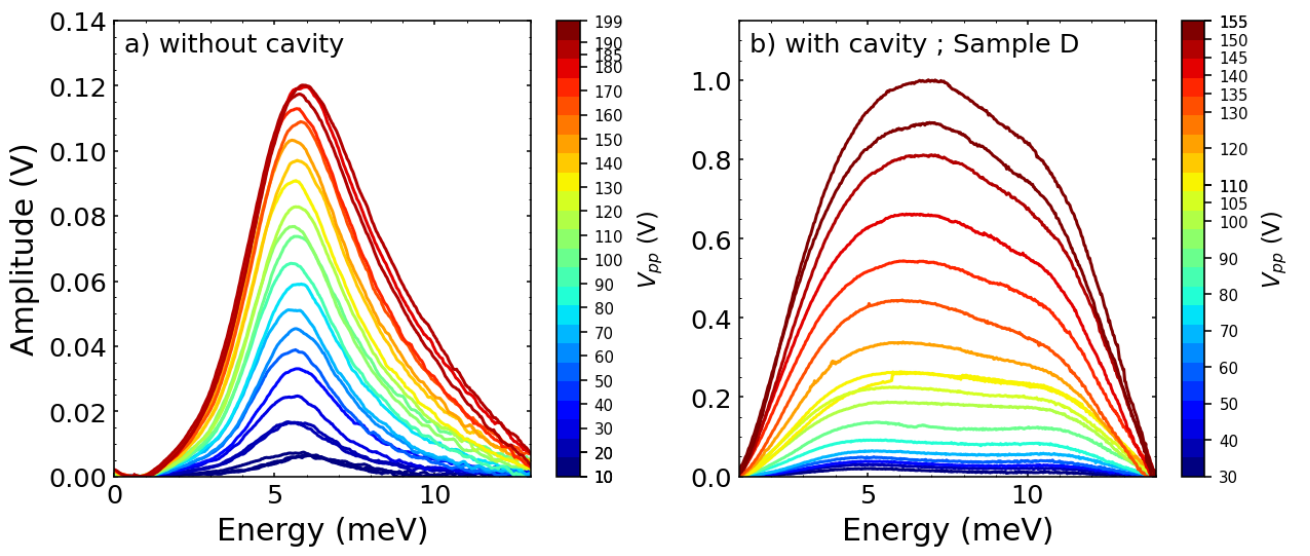

FIG. S.13. Waterfall plot of the emission spectra obtained at a fixed magnetic field of 0.9 T while increasing the injected pulse amplitude, for sample D measured (a) without and (b) with the cavity.

### Strong Coupling Evidence and Quantitative Analysis

A clear anticrossing between two resonances constitutes the unambiguous signature of the strong coupling regime. In our system, two anticrossings are indeed observed, corresponding to the first and second Fabry–Perot cavity modes (see Fig. S.14). The first cavity mode exhibits a distinct splitting at zero detuning, with two peaks centered at approximately 2.4 meV and 3.7 meV, yielding a coupling ratio $g/\omega = 21\%$. For the second cavity mode, the two branches are less resolved at zero detuning, with overlapping peaks at around 10 and 11.5 meV. Nevertheless, the anticrossing between the two resonances remains clearly visible.

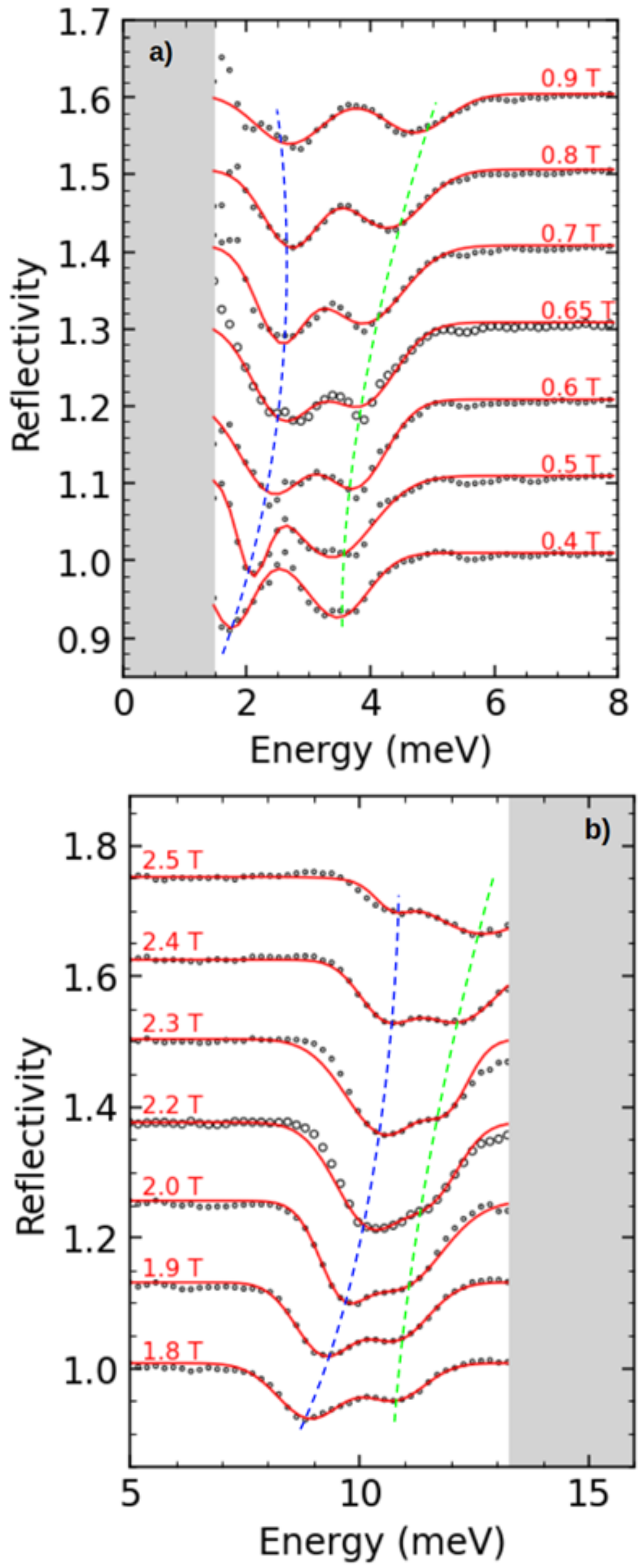

FIG. S.14. Waterfall plot of the magneto-reflectivity spectra obtained for different magnetic field strengths, around the first (a) and the second (b) anti-crossing, in sample A. The spectra corresponding to the zero detuning are plotted with empty black circles. The red lines are double Gaussian fitting to highlight the positions of the lower and upper polariton branches. The blue and green dashed lines are guide for the eye to emphasize the avoided crossings.

### Sample B characterization

Figure S.15a shows the emission spectra of sample B as the magnetic field increases. The first cavity mode (3–4 meV) appears at low fields and gradually shifts to higher energies. Around 0.6 T, a second mode (10 meV) becomes visible, while the emission progressively transfers from the first to the second. At 0.9 T (bold spectrum), both modes coexist, with the second still weaker. This coexistence indicates a smooth transition between the two polaritonic branches as the resonance condition evolves. Figure S.15b shows the spectrum at 0.9 T without the cavity; the dashed lines mark the energies of the two modes (UP1 and UP2), matching the regions highlighted in Fig. 3d of the main text.

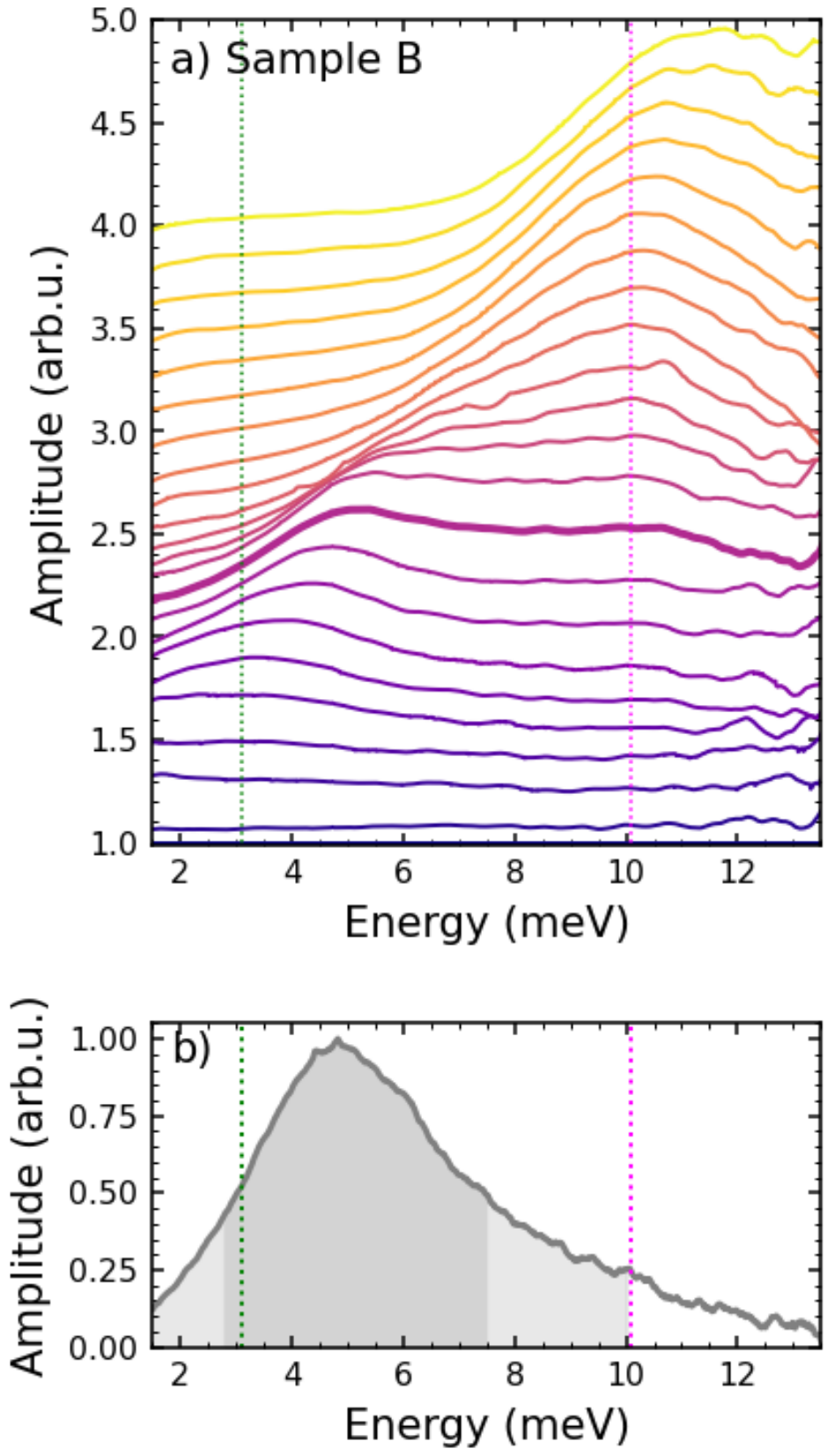

FIG. S.15. (a) Waterfall plot of raw emission spectra from sample B with increasing magnetic field (0–1.9 T). At 0.9 T (bold curve), two peaks are visible, corresponding to the first and second cavity modes. (b) Bare CR emission obtained without the cavity at a magnetic field of 0.9 T. In both panels, the green (resp. pink) dotted line highlights the UP1 (resp. UP2) energy position.